# Statistical properties of microcracking in polyurethane foams under tensile test, influence of temperature and density

S.Deschanel [a,b], L.Vanel [a], G.Vigier [b], N.Godin [b], S.Ciliberto [a]

[a] Laboratoire de physique, CNRS UMR 5672, Ecole Normale Supérieure de Lyon, 46 allée d'Italie, 69364 Lyon Cedex 07, France

[b] Groupe d'Etudes de Métallurgie Physique et de Physique des Matériaux, INSA de Lyon, 20 Av. Albert Einstein, 69621 Villeurbanne Cedex, France

**Abstract**

We report tensile failure experiments on polyurethane (PU) foams. Experiments have been performed by imposing a constant strain rate. We work on heterogeneous materials for whom the failure does not occur suddenly and can develop as a multistep process through a succession of microcracks that end at pores. The acoustic energy and the waiting times between acoustic events follow power-law distributions. This remains true while the foam density is varied. However, experiments at low temperatures (PU foams more brittle) have not yielded power-laws for the waiting times. The cumulative acoustic energy has no power law divergence at the proximity of the failure point which is qualitatively in agreement with other experiments done at imposed strain. We notice a plateau in cumulative acoustic energy that seems to occur when a single crack starts to propagate.

**1. Introduction**

Damage mechanisms up to rupture in heterogeneous materials have recently received a lot of attention in the scientific community [1-4]. Improvements in acoustic emission (A.E) technique have permitted the monitoring in real time of the gradual damage of stressed materials [5], and in particular to spatially localize the A.E and even identify the rupture mechanisms [6-12]. Due to microcrack arrest at defects, the failure of heterogeneous materials may develop as a multistep process through a succession of local events leading to diffuse damage. As a result, fracture in an heterogeneous material can be often described as a clustering of microcracks. Recent theoretical works have considered this rupture process as a second order phase transition [13,14] where the final failure is the critical point of this phase transition. This approach gives a new perspective and method for anticipating failure. Lately, experimental works have shown the relevance of this theoretical approach and in particular the possibility to have a better prediction of failure time in the case of fiber composites [15-23]. One important observation is that the damage rate in a brittle and heterogeneous material, for instance estimated in terms of cumulative acoustic energy released, presents a critical divergence close to the failure time. In addition, acoustic emissions present power law behaviors showing they have no characteristic scales of energy and time. In order to probe the generality of these observations, it is of importance to understand whether they depend either on the type of heterogeneous structure or on the mechanical behavior (brittle or ductile) of the material.

Damage growth in composites involves many mechanisms including matrix cracking, fiber-matrix interface debonding, fiber rupture and delamination [24,25]. For an easier understanding of damage growth in heterogeneous materials, we choose to study a simpler and well controlled material such as vitreous polymer foams, which are frequently used in different technological applications. They are composed of a single constituent, the degree of heterogeneity (amount of voids) can be adjusted through the elaboration of the materials and their mechanical properties change with temperature from brittle to ductile behavior. Specifically, we have used polyurethane foams recognized as ductile at room temperature.

We record acoustic emission activity emitted by a sample originating from damage to the material. The damage associated to an acoustic event can be the appearance either of a single crack or of several cracks which cannot be resolved in time. The A.E, which are elastic waves emitted within a material, can be detected by piezoelectric sensors fixed onto the surface of the specimen. We want to probe the influence on the A.E statistical properties of test conditions, physical properties or morphology of the material. In this paper, we change morphology by varying material's porosity and we change the mechanical behavior by varying temperature. This will allow us to study the generality of the statistical properties of rupture in heterogeneous materials.

**2. Experimental procedure**

2.1. Materials

The samples of polyurethane foams (PU) are obtained by first mixing two polyols and catalysts, with a stochiometric ratio (polyethers are the more commonly used polyols). Then a silicon based surface active agent and water are added [26] and the whole is stirred for 2 minutes using a mechanical agitator at 600 rpm. A polyisocyanate (diphenyl methane diisocyanate) is then added and the components are mixed for 20s at 1200 rpm at which point they are poured into an open cylindrical mold and allowed to cure. The basic reactions involved in the production of the polyurethane foams are often referred to as the blowing and gelling reactions. The reaction products of the blowing reaction between the diisocyanate and water are carbon dioxide which foams the reacting mixture and a distributed amine. The amine produced in this reaction reacts with additional isocyanate to produce rigid urea groups which, when of sufficient size and concentration, phase separate into urea rich domains (hard segment domains) primarily due to hydrogen bonding with additional urea groups. The hard segment domains (structure, order, concentration) play a very important role on the final structure, morphology, and properties of the foam [27]. The reaction product between the isocyanate and multifunctional polyol is a urethane group which links the urea groups to the ether soft segments [28].

The foam expansion is executed in three minutes just before the gelation (polymerization/cross-linking). Depending on the density required, different amounts of water are added. In fact, water concentration controls the expansion of the foam. The different relative densities investigated in this study vary from 0.4 to 0.9. The relative density is defined as the ratio $\rho^*/\rho_s$ where $\rho^*$ is the apparent density of the foam and $\rho_s$ the PU density. The glass transition temperature, measured by Differential

Scanning Calorimetry, is at 75°C. This later is measured constant over all the studied density range. This transition is spread over about twenty degrees, which is an advantage for experiments in that the variation in properties is continuous and non sudden with temperature change. The dynamic mechanical behavior of the PU foam indicates a secondary relaxation to be around -75°C at a frequency of 1Hz.

The samples morphology is investigated by scanning electron microscopy (SEM) and X-ray tomography. In Fig. 1, four different relative densities of closed cell PU foams are represented, these pictures are 2D extracted tomographic slices. The mean diameter of the PU foams vary from 40µm to 100µm respectively for relative densities from 0.9 to 0.4. Their microstructure change with density. The pore size distribution is sharper for high density and when the foam has a higher porosity the pore sizes are more dispersed (Fig. 2).

The relative density of the specimens are calculated by measuring the weight and the volume of each specimen. They are also verified using SEM or tomographic observations by calculating the ratio pores to polymer phase.

2.2 Mechanical testing

Tensile tests are performed on PU foams (from 0.4 to 0.67 relative density) at room temperature using an MTS hydraulic machine in displacement control at a constant crosshead speed of 0.3 mm.min$^{-1}$ (strain rate of $1.10-4s^{-1}$). The specimens are machined for mechanical tests with dimensions : $3\times10\times50mm^3$.

Tensile tests are also performed at different temperatures (-10°C, -30°C and -65°C) on PU foams of 0.58 relative density. The load and strain (determined from the cross-head displacement) are recorded up to the macroscopic fracture.

2.3 Acoustic emission (A.E)

Acoustic emission is continuously monitored during the tests using a Mistras 2001 data acquisition system of European Physical Acoustic (EPA) with a 8 MHz sample rate and a 40 dB pre-amplification, the bandwidth being 50 kHz – 1.2 MHz. Our measurements are achieved with two resonant R15 EPA sensors (peak of resonance at 150kHz) coupled to the material with silicon grease. The sensors are held in contact thanks to a silicone tape and placed at a nominal distance of 40mm between their respective centers.

When recording A.E signals, the user fixes a detection threshold (28dB to 33dB depending on tests conditions), below which no signal is to be recorded. Several parameters determined from the waveforms are recorded for each acoustic event : maximum amplitude, energy, rise time, counts (number of threshold crossing) and duration. The amplitude distribution covers the range 0-100dB (0dB corresponding to 1µV at the transducers output) and the energy is obtained by integration of the squared signal.

A pencil lead break procedure [29] is used to simulate A.E signals during calibration of each test. These preliminary measurements (where a repeatable acoustic wave can be generated) allow us to set up the acquisition parameters for our materials : peak definition time = 200µs, hit definition time = 400µs, hit lock time = 800µs. At the same time, we measure the attenuation and the mean wave speed for each sample (the difference in arrival are deduced from the first peaks detected). This velocity determination procedure was repeated several times and at different positions between the sensors to assure the accuracy of the measured wave speed. At room temperature, the velocity for the PU foams is in a range of 1000m.s$^{-1}$ to 1700m.s$^{-1}$ (depending on the relative density from 0.4 to 0.9 respectively). The sound velocity does not seem to change significantly with damage. For a PU foam of relative density 0.58, the wave speed at room temperature, $-10°C$, $-30°C$ and $-65°C$ are respectively 1250m.s$^{-1}$, 1300m.s$^{-1}$, 1400m.s$^{-1}$ and 1800m.s$^{-1}$. When the sensors are 35mm apart, the signal attenuation is 15dB for the denser foams and 40dB for the less dense ; this attenuation follows an exponential law as a function of the distance to the sensor. Otherwise, when the temperature is decreased, the attenuation is less and less important.

For each experiment, we monitor the mechanical stress versus strain and we try to correlate this evolution to the number of A.E events. In addition, the position of microcracks created under stress is estimated along the specimen length. Maps of acoustic emission signals location (positioned on the specimen) at various loading stages has already been shown [17,30]. In our set-up, as we use only two sensors, we have access to linear location : the knowledge of the wave speed in the material and the difference in arrival times at each sensor of a given wave is not sufficient to determine the precise location of the A.E. However, we can get an idea of the typical distance from the sensors of the A.E assuming it occurred on the median line joining the centers of the two sensors. Besides, only a certain amount of events can be situated along the specimen : only those whose intensity is big enough to reach both sensors can be localized. Then, we will differentiate all the A.E signals collected from the localized signals.

Furthermore, we consider the energy released ($\varepsilon$) by a damage event and the time intervals ($\delta t$) between two events as the main characteristics describing the fracturing process. Due to attenuation, the energy measured by each sensor depends on the distance of the event from the sensors. In order to obtain an energy independent of the location where the event occurred, we determine the energy released $\varepsilon$ as $\varepsilon = \sqrt{\varepsilon_1 . \varepsilon_2}$ with $\varepsilon_1$ and $\varepsilon_2$ the energies received by the two sensors. We examine the probability distribution $N(\varepsilon)$ of the energy $\varepsilon$ and the probability distribution $N(\delta t)$ of the times $\delta t$ between two consecutive events.

2.4 X-ray tomography

To illustrate the microscopic cracking event, we combine monitoring of the acoustic emission during a mechanical test and a X-ray tomography technique. X-ray computed microtomography allows the study of deformation mechanisms of foams when coupled with in situ loading tests [31]. According to the Beer Lambert law, each element in the recorded projection corresponds to a line integral of the attenuation coefficient along the beam path. The resulting image is a superimposed information of a volume in a 2D plane. To get 3D images, a larger number of radiographs are recorded while rotating the sample between 0° and 180°. Our experiment was carried out at the BM05 beamline of the European synchrotron radiation facility (ESRF) in Grenoble, France. In situ experiments are performed on specimens thinner at the center, using a special rig designed for this purpose [31]. Further details on these experiments will be described elsewhere [32].

**3. Experimental results**

The evolution of the stress together with the cumulative number of A.E events are recorded versus the strain during each tensile test. The cumulative number of A.E events is a global measure which enables the estimation of the damage level during loading.

<u>3.1 Typical behavior of an heterogeneous material under stress at room temperature</u>

Let us take the example of a tensile test on a PU foam sample whose relative density is 0.67 (Young modulus E=690MPa). The stress/strain curve is represented with the A.E activity in Fig 3a. We ascertain that an A.E activity is detected, yet only a certain amount of events can be localized. Both the total amount of A.E signals collected (▲) and the localized signals (o) are shown in Fig 3a. Their evolution is similar : there is a quiet phase below 4% of strain with very few signals detected ; then the emission rate increases significantly until final rupture characterizing the damage response of the material under load. In fact, there is very little acoustic activity during the elastic strain while this activity increases during the plastic strain as deformation becomes irreversible. In the following of this paper, only events localized between the sensors will be used for analysis. In that way, we ensure that signals effectively come from the material itself. Linear location along the specimen is represented in Fig 3b. We observe that the final failure at time $\tau$ corresponds to an important concentration of events.

Taking into account the number of events is not sufficient to understand the failure process ; studying the energy gives additional information about the extent of damage. Indeed, we might think that each A.E burst corresponds to one microfracture but it can not be asserted that each microfracture corresponds to the breaking of a single wall between adjacent pores. In fact, several walls might break in cascade. Moreover it is likely that the strength of each wall is different and that a wall can break progressively. In that way, energy seems to be an adequate characteristic parameter of the system to characterize the divergence we observe with the number of events near the fracture point (Fig. 3a). First, we characterize the statistical properties of the AE events themselves, looking at the probability distribution N($\varepsilon$) of the energy released and at the distribution N($\delta$t) of the time intervals between two consecutive A.E events ($\delta$t). $\delta$t gives information about the dynamic of the failure process. A typical probability distribution N($\varepsilon$) is plotted as a function of $\varepsilon$ in log-log scale in Fig 4a (for the case of one tensile test on a PU foam sample). The upper and lower energy limits are, respectively, the strongest event recorded and the threshold (the minimum energy value the acquisition system can detect). A power-law is revealed spanning through almost four decades : $N(\varepsilon) \propto \varepsilon^{-\alpha}$. Likewise, N($\delta$t) has a power-law dependence on time : $N(\delta t) \propto \delta t^{-\beta}$ over almost five decades (see Fig. 4b). The mean exponents found are $\alpha$=1.47±0.09 and $\beta$= 1.28±0.08 for ten samples of PU foams with a relative density of 0.67.

The power-law behavior indicates that overall the system does not have a characteristic scale of energy or time. These results are consistent with previous experimented works on various heterogeneous materials as in paper [33] and in fibrous composite materials [17].

Now, we turn to an analysis of the divergence of AE activity observed close to rupture for the PU foam specimens loaded in tensile tests at constant strain rate. Guarino et al. [16-18,20] showed that the cumulative energy was described by a power-law near the breaking time $\tau$, as a function of the reduced control parameter $(\tau-t)/\tau$, in the case of creep tests. As $\varepsilon$ is an intermittent variable, they preferred to use as susceptibility the cumulative energy $E_{cum}$. Following this approach, we study the cumulative energy $E_{cum}$ of the localized events, normalized to $E_{max}$, as a function of the reduced parameter $(\tau-t)/\tau$, $E_{max}$ being the total energy. The normalized cumulative number of events and the normalized cumulative energy emitted during tensile tests of PU foams of relative density 0.67 are plotted in Fig. 5 versus $(\tau-t)/\tau$ in log-log scale ; the stars and circles are the average for 8 samples. Qualitatively, we notice that the evolution of both curves is almost similar. There is a first part where the cumulative energy and the number of events increase slowly : during 90% of the total time, only 10% of the energy has been released and only 10% of the total number of events has been recorded. Then both of them increase up to a point where there is a plateau : this last stage appears very late, at only 0.04% of the time before failure. But, there is no critical divergence observed in that study. We emphasize that the range of reduced time is very large and permits to follow the dynamics 3 to 4 orders of magnitude closer to the failure point than in [16-18,20]. The inset of the Fig. 5 shows the mean of the logarithmic values of ε as a function of the reduced time : this gives an idea of the typical energy released by an event. The events have higher energy when the time is close to failure, that is why the two curves should differ mainly in the vicinity of the fracture point.

3.2 Influence of the porosity

Tensile tests are carried out at room temperature for different densities of PU foams to determine the influence of the material heterogeneity on the critical exponents of the power-laws. The different stress strain curves for relative densities $\rho^*/\rho_s$=0.4, 0.58, 0.63 and 0.67 are plotted in Fig. 6. with their respective acoustic activity. We observe an increasing evolution of the Young modulus and the yield strength of the different foams with the relative density. For the different densities studied, the Young modulus and maximum stress are summarized in table 1 :

| Relative density ($\rho^*/\rho_s$) | Young Modulus (MPa) | Maximum Stress (MPa) |
|---|---|---|
| 0.4 | 250 | 6 |
| 0.58 | 550 | 12 |
| 0.63 | 650 | 15.5 |
| 0.67 | 690 | 17 |

**Tab. 1** Mechanical characteristics for different PU foam densities

The ratio, foam's Young modulus E* over dense polyurethane's Young Modulus $E_s$ (E*/$E_s$, with $E_s$=1380) may be related to the relative density for tests at room temperature. We find that the evolution of E*/$E_s$ is well described by a polynomial function of the second degree which allows to determine $E^*/E_s$ as a function of $\rho^*/\rho_s$ : $E^*/E_s = 0.89(\rho^*/\rho_s)^2 + 0.17(\rho^*/\rho_s)$. This evolution is in agreement with the model established by Gibson and Ashby for closed cell foams [34]. When looking at the acoustic activity, the evolution of the cumulative number of events is qualitatively identical for each density : very few events at the beginning and an increasing number during the plastic plateau.

Fig. 7. a) and b) represent the probability distributions of $\varepsilon$ and $\delta t$ for the different relative densities. The exponents of the scaling law $N(\varepsilon) \propto \varepsilon^{-\alpha}$ (Fig. 7. a) are found to be close to each other : the slopes are almost parallels. The value of the exponent $\alpha$ is 1.42±0.15 when averaged over the different densities. In the case of the scaling law $N(\delta t) \propto \delta t^{-\beta}$ (Fig. 7. b), the slopes of the different fits are also almost identical ; the exponent $\beta$ has a mean value of 1.25±0.14. We do not observe any systematic variation of $\alpha$ and $\beta$ with material density.

The evolution of the cumulative energy for the different relative densities is plotted in Fig. 8. At the beginning, the cumulative energy for all the different foams follows approximately the same evolution. Then, there is a small difference in the evolution of each curve : for the foams of higher densities ($\rho^*/\rho_s$= 0.63 and 0.67), 90% of the acoustic activity is recorded during the last 10% of the time while for the less dense foams, 75% and 70% (respectively for relative densities 0.4 and 0.58) are to be recorded during the last 10% of the time. Then, the respective cumulative energy increase up to 0.2% of time before failure from where they saturate together in a slope close to zero (in log-log scale). Although such a behavior has already been observed at imposed strain in other materials [17], the saturation occurs here for much smaller values of the reduced time. In fact, for reduced time values typically between $10^{-1}$ and $10^{-3}$ we do observe a significant variation of the cumulated energy, sometimes close to a power law, contrary to observations reported in [17].

3.3 Influence of the temperature

Tensile tests at different temperatures (from room temperature down to –65°C) have been performed on PU foams of relative density 0.58. The foam becomes increasingly brittle with temperature decreasing, as shown in Fig.9 where the mechanical behavior is plotted with the cumulative number of A.E events. The Young modulus is increasing with temperature decreasing, likewise the maximum stress which almost double between room temperature and $-65°C$. The Young modulus and maximum stress are summuarized in the table 2 :

| Temperature (°C) | Young Modulus (MPa) | Maximum Stress (MPa) |
|---|---|---|
| 26 °C | 550 | 12 |
| -10 °C | 620 | 15.5 |
| -30 °C | 700 | 18.5 |
| -65 °C | 800 | 22.4 |

**Tab. 2** Mechanical characteristics for a PU foam (relative density 0.58) at different temperatures

We also notice that the plastic plateau disappears with decreasing temperature, and that the failure strain is less important, the material getting more and more brittle.

The behavior in A.E at room temperature and at $-10°C$ are almost identical : acoustic activity begins late when the material is already in the plastic plateau stage and a divergence of the number of A.E events occurs at the end (Fig. 9). The microcracks nucleate, concentrate and coalesce at the end of the test, producing the final failure (Fig. 10a). Final failure takes place at the coalescence sites. On the contrary, for the tensile tests effectuated at $-30°C$ and $-65°C$ the A.E activity starts at the very beginning, indicating the early occurrence of damage (Fig. 9). The number of events rises gradually as the load increases. The signals are shown to emanate from sources gathered in a specific location from the start and no clustering of acoustic events are observed before failure (Fig. 10b). Hence, we have observed a difference in behavior for two groups of temperatures : room temperature and $-10°C$ compared to $-30°C$ and $-65°C$.

The probability distribution N($\epsilon$) of the energy $\epsilon$ reveals a power-law at every temperature (Fig. 11a). In addition, the value of the critical exponent $\alpha$ of this law does not seem to depend on the intrinsic properties of the material. Indeed, considering the measurement incertitude, the exponent shows little variation around $\alpha$=1.4. This is not the case for the probability distribution N($\delta$t) (Fig.11b). A power-law is obtained in the experiments at room temperature and $-10°C$ with a value of β=1.38±0.1 but not at $-30°C$ and $-65°C$. At this lower temperatures, there is no plateau where the load is quasi constant.

Likewise, the difference in behavior of the two groups of temperatures is observed with the evolution of the cumulative energy (Fig 12). For the tests at room temperature and –10°C, the evolution of the cumulative energy is identical to the one already explained at room temperature. On the other hand, the cumulative energy at –30°C and –65°C has a different evolution : it increases sharply at the beginning and afterward stays quasi constant (log-log scale speaking). During 80% of the total time, only 10% of the energy has been released for the foams tested at room temperature and –10°C while at this same time, the cumulative energy is already at 80% of the total for the tests at –30°C and 92% for the tests at –65°C. From the point 0.1% of reduced time $(\tau-t)/\tau$, the curves are similar : there is a kind of divergence but on a very small logarithmic scale of $E_{cum}/E_{max}$.

## 4. Discussion

4.1 Relation microstructure / A.E : source mechanisms identification

We analyze the deformation mechanisms during in situ tensile tests by means of X-ray microtomography combined with acoustic emission monitoring. Further details on these experiments will be described elsewhere [32]. These tests are performed on PU foams of relative density ρ*/ρ$_s$= 0.9, for which the strain to rupture reaches 30% before failure. The qualitative results obtained are useful for a better understanding of our previous observations. We notice that the A.E signals may come from different sources during the test. For the PU foam tested at room temperature, the lack of A.E signals at the beginning of the loading may be due to the deformation of the material coming without any wall breaking (verified during in situ experiments combined with A.E monitoring [32]). Then, above a certain degree of strain, walls between cavities begin to collapse and microcracks start to grow but stops as soon as it encounters a pore. In fact, A.E signals represent wall breaking between adjacent pores (Fig. 13, 16% of strain). Afterwards, crack propagation is more likely to happen towards the end of the test (Fig. 13, 22% of strain). The separated events lead to a more important one : the microcracks coalesce and culminate in the catastrophic rise of a global crack implying the material's fracture. Actually, the concentration of microfractures may be a good indicator that the sample is approaching failure.

4.2 Typical analysis

We noticed that the A.E energy released and the elapsed time between consecutive events have an invariant power-law distribution in the case of the tensile tests effectuated at room temperature. In that way, we present experimental evidence for scale invariance in microfracturing processes via the acoustic emission.

Now, we attempt to interpret the different stages in the evolution of the cumulative energy using tomography experiments. The first 90% of the time where the cumulative energy increases gradually (Fig. 5) may corresponds to the part where the microcracks appear at different locations (wall breaking between adjacent pores here and there), the events are not correlated. Afterwards, we might think that the events are more and more correlated, the microcracks being close to the coalescence. Finally, the plateau may stand for the appearance of the final crack in a very short time, although there is no clear power-law for the cumulative energy $E_{cum}$. Likewise, Salminen et al [33] have shown on experiments on paper that there is no clear sign of a "critical point". The picture we give here is quite general. It is similar to experimental observations in fibrous composite materials [17] and in granite samples [35,36] where three stages have been distinguished : at the beginning, microfractures are roughly uniformly distributed, afterwards they begin to concentrate and then they grow to form a single crack.

4.3 Influence of the porosity

We have shown that the critical exponents of the power-laws $N(\varepsilon) \propto \varepsilon^{-\alpha}$ and $N(\delta t) \propto \delta t^{-\beta}$ do not vary substantially with porosity. Besides, the evolution of the cumulative energy is similar for each foam. This highlight the presence of scale invariance on microfracturing process as it has already been observed in many situations : the fracture of granite [35,36], the acoustic emission from volcanoes [37], chemically induced fracture [38], the fracture of plaster samples cracked by piercing through them [39], the explosion of a spherical tank [15], the fracture of fibrous composite materials [17,18,20] and of cellular glass [40]. The power-law we observe have identical exponents to those measured in other setups including computer simulations for models which are based on Self Organized Criticality. The value of α we obtain is close to the one given in [17,40] and [37,38,41-43] where α=1.5 for creep experiments. As well, the value of the exponent β is not too different from [37,40,41].

4.4 Influence of the temperature

We pointed out two different behaviors for two groups of temperatures : room temperature and -10°C versus -30°C and -65°C. For the foams studied at room temperature and -10°C, there is no acoustic activity at the beginning because the pores lengthen and do not break (verified during in situ experiments with X-ray tomographic technique at room temperature). Then, walls break, cracks grow and percolate to failure. On the other hand, for the foams studied at lower temperatures, A.E events are recorded as soon as the loading starts : as the material is more brittle, the heterogeneities can create much more stress concentration than in the case where plastic deformation would occur. This explains that it is possible to start breaking the material even for small applied load. At $-65°C$, the foam is even more brittle and material deterioration occurs without any higher concentration of microfractures at the end of the test (Fig.10.b) : it may corresponds to a single crack growing from almost the beginning of the test. For tests realized at $-30°C$ and $-65°C$ we noticed that there is no power-law for the interarrival times δt distribution although the energy ε is power-law distributed at any temperature. A hypothesis that could be formulated is that δt is power-law distributed for tensile tests only when stress remains quasi constant during acoustic activity i.e during damage. This is the case for tensile tests at room temperature and -10°C since most of the acoustic activity occurs during the plateau of the stress-strain curve. On the contrary, for tests realized at $-30°C$ and $-65°C$, acoustic activity occurs in the linear part of the stress/strain load curve and no power law is observed.

The distribution of the energy seems little influenced by the intrinsic properties of the material. Conversely, the brittle or ductile nature of the material seems to affect the distribution of interarrival time. We infer from these observations that the distribution of δt is sensitive to the details of the failure processes. In that sense, we might say the real control parameter of the failure process is time. A similar conclusion has been reached in [20] for other fibrous materials. On the other hand, we show that the distribution of ε is a general feature of fracture in heterogeneous materials and is not linked to a specific propagation mode. Now, the dissimilarities in the progress of the cumulative energy may be explained by the fact that at lower temperatures, the samples being more brittle, the cracks propagation is enhanced and appears earlier. That is why the plateau observed in Fig. 12 appears earlier in time than for the tests at higher temperatures.

## 5. Conclusions

Failure of polyurethane foams shows some features associated to second order phase transition at the critical point. The behavior of the AE event energies and the AE event intervals follows power-law-like statistics with several order of magnitude of scaling. The exponents of these laws remains roughly the same for all the foam densities studied. The energy is power-law distributed at any temperature while the waiting time distributions do not show power law when the material is more brittle (at lower temperatures). The in situ tensile tests with X-ray tomography technique allows us to understand the source mechanisms of AE for tests at room temperature. At the beginning, there is no acoustic activity and the pores elongates, then the walls between pores break and crack growth predominates while the number of AE diverges. Our experimental measurements of AE show that there is no clear sign of critical divergence of the cumulative energy near the fracture point. We show that the observed plateau seems to be well correlated to the propagation of a single crack, especially at low temperature where a crack starts to propagate very early in the test. The observation of a plateau in cumulative energy seems to be consistent with the results of [17,20], where divergence has been observed only for tensile tests performed at controlled stress rate. However, we note that in the same range of reduced time than in [17], the cumulative energy we measured do vary significantly, almost as a power law in some cases. We are now performing creep tests on these materials. Work is in progress and the results will be the subject of another report.

**Fig. 1** 2D extracted tomographic slices of PU foam of different relative densities

**Fig. 2** Particle size analysis of PU foams of different porosities

**Fig. 3** a)Stress/strain curve with acoustic emission activity b) Linear location of AE signals along the specimen (stressed in length direction) during tensile test ($\dot{\varepsilon}=1.10^{-4}$ s$^{-1}$) for a PU foam of relative density 0.67

**Fig. 4** Probability distributions of (a) ε and (b) δt obtained for one tensile test at constant strain rate on a PU foam sample of relative density 0.67 ; the solid lines are power fits which exponents are respectively α=1.5 and β=1.3.

**Fig. 5** Cumulative normalized energy $E_{cum}/E_{max}$ and number of events versus reduced time (τ-t)/τ. The circles and stars are the average for 8 samples of PU foams of relative density 0.67. The inset shows the mean of log(ε) versus reduced time.

**Fig. 6** Stress/strain curves with acoustic emission activities during tensile tests for different relative densities of PU foams (from 0.4 to 0.67).

**Fig. 7** Probability distributions of (a) ε and (b) δt obtained for PU foams of relative density from 0.4 to 0.67 during tensile tests (room temperature); the solid lines are power fits.

**Fig. 8** Cumulative normalized energy $E_{cum}/E_{max}$ versus reduced time (τ-t)/τ for the relative densities from 0.4 to 0.67.

**Fig. 9** Stress/strain curves with acoustic emission activity during tensile tests for PU foams (ρ*/ρs = 0.58) at different temperatures.

**Fig. 10** Linear location of AE events along the specimen and stress versus time at (a) -10°C and (b) -30°C (PU foam, ρ*/ρs = 0.58).

**Fig. 11** Probability distributions of (a) ε and (b) δt obtained for PU foams of relative density 0.58 during tensile test at –10°C and at –30°C; the solid lines are power fits.

**Fig. 12** Cumulative normalized energy versus reduced time for different temperatures (tensile tests, PU foams of relative density 0.58).

**Fig. 13** 2D extracted tomographic slices showing different strain levels during a tensile test on a PU foam of relative density 0.9 at room temperature.

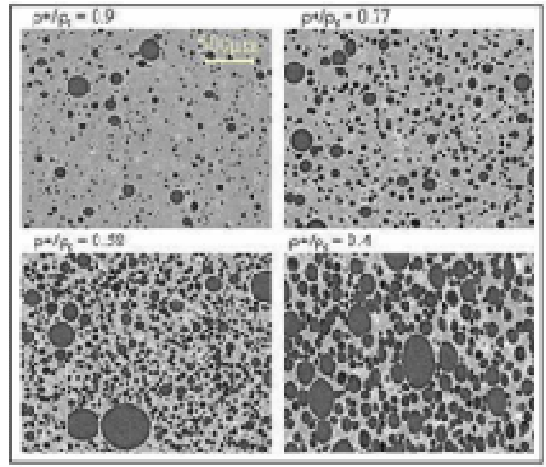

**Figure 1**

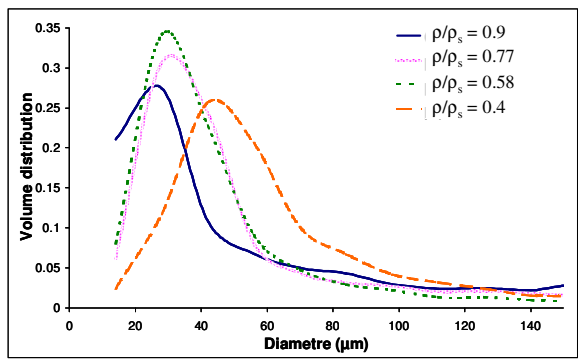

**Figure 2**

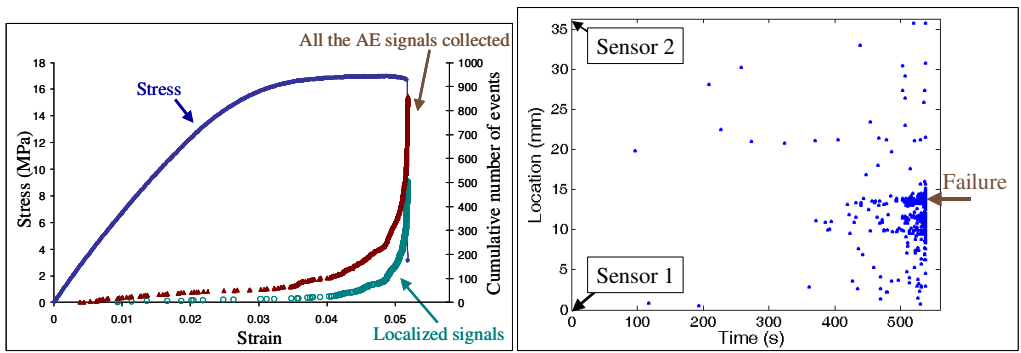

**Figure 3**

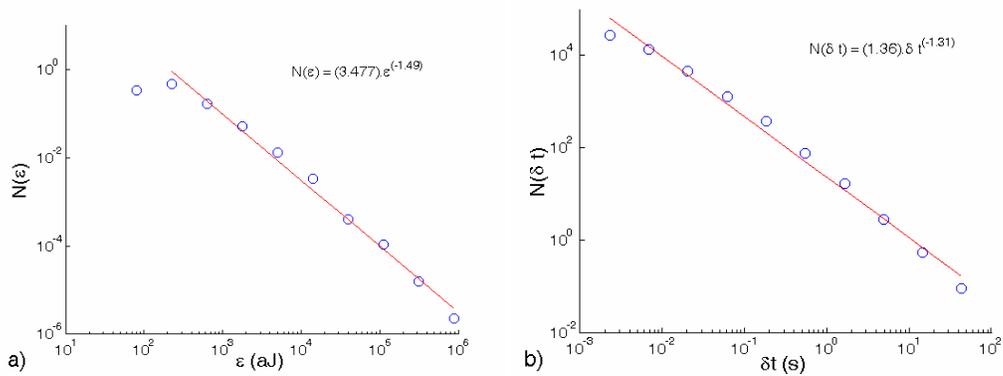

**Figure 4**

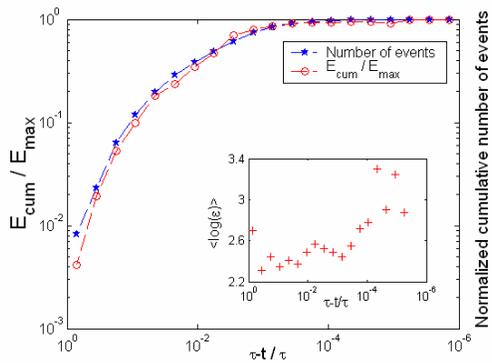

**Figure 5**

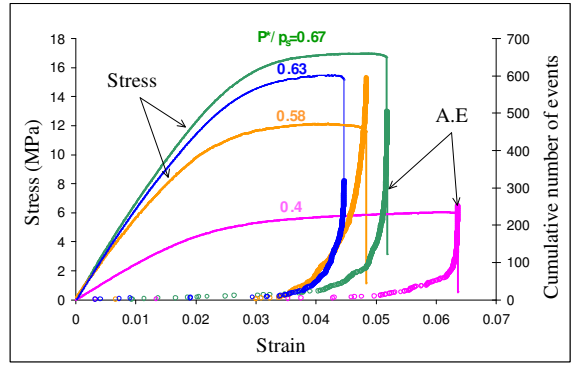

**Figure 6**

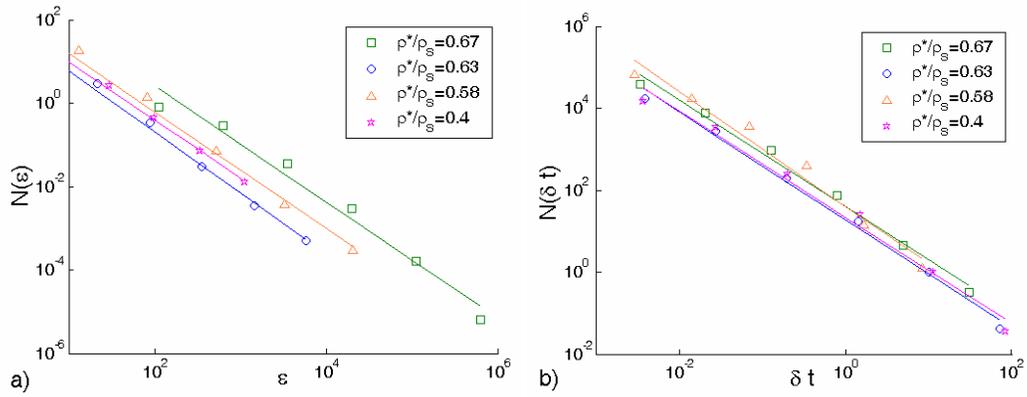

**Figure 7**

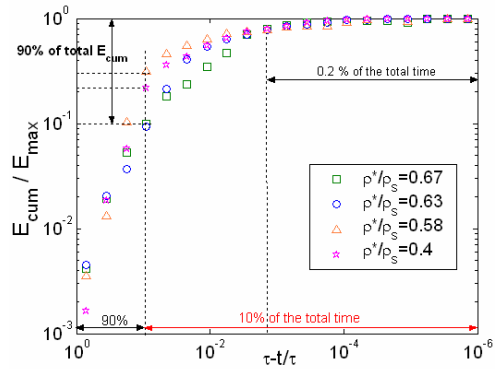

**Figure 8**

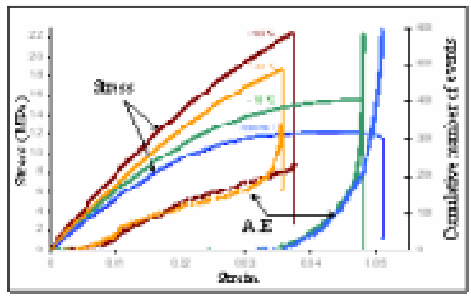

**Figure 9**

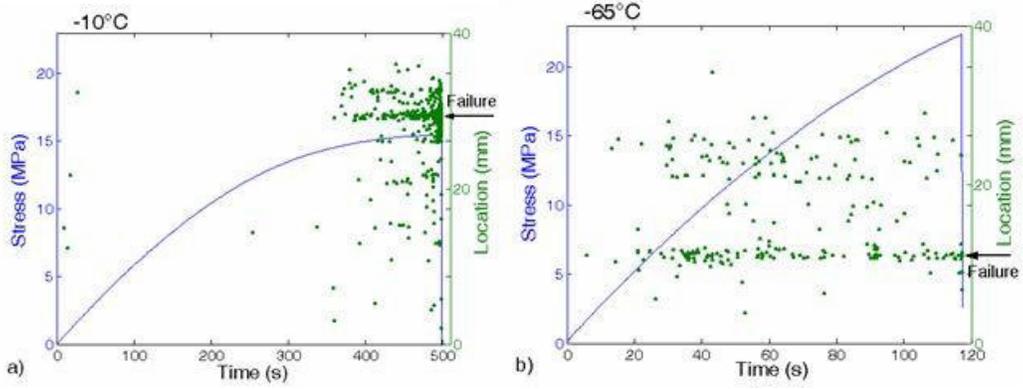

**Figure 10**

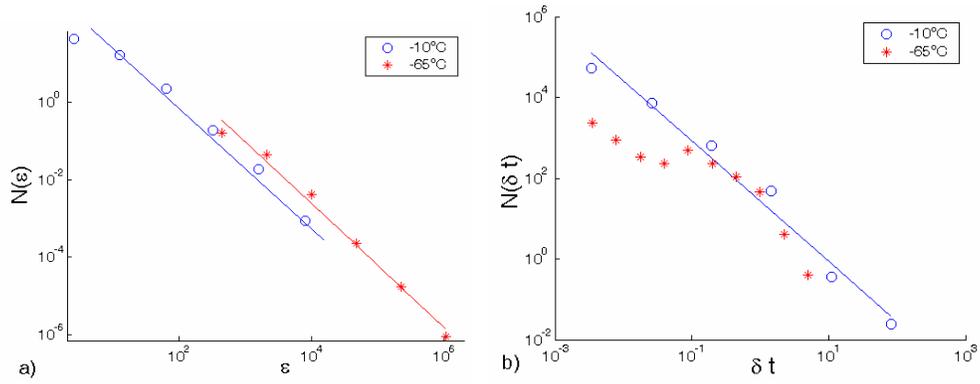

**Figure 11**

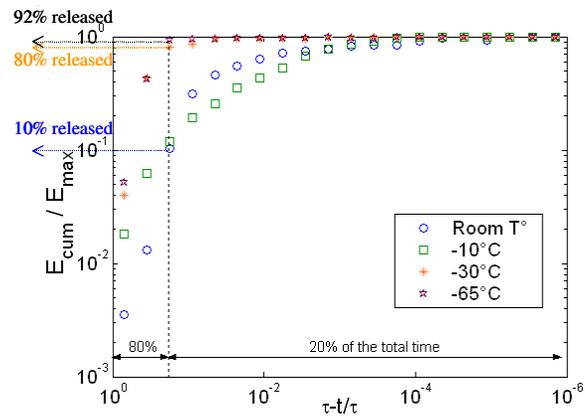

**Figure 12**

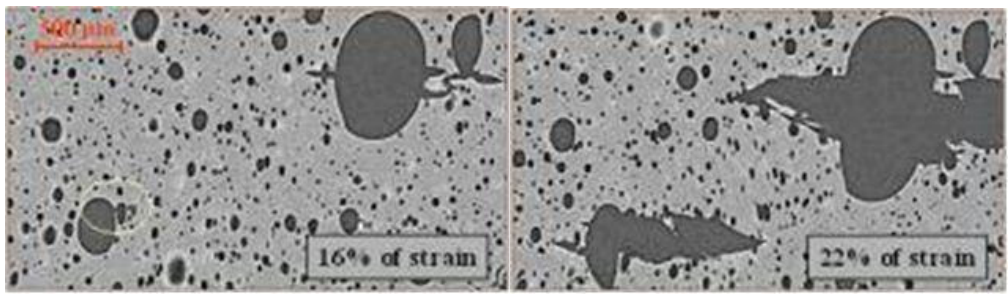

**Figure 13**

22